\documentclass[conference]{IEEEtran}
\IEEEoverridecommandlockouts
\usepackage[noadjust]{cite}
\usepackage{amsmath}
\usepackage{array}
\usepackage{mdwmath}
\usepackage{amssymb}
\usepackage{mathtools}
\usepackage{algorithm}
\usepackage[noend]{algpseudocode}
\usepackage{eqparbox}
\usepackage{stfloats}
\usepackage{url}
\usepackage[pdftex]{graphicx}
\graphicspath{{../pdf/}{../jpeg/}}
\DeclareGraphicsExtensions{.pdf,.jpeg,.png}
\usepackage{epstopdf}
\usepackage{amsfonts}

\usepackage{color}

\usepackage{tikz}
\usepackage{pgfplots}
\usepackage{subcaption}
\usetikzlibrary{spy}

\begin{document}

\title{\huge Hybrid Digital-Wave Domain Channel Estimator for Stacked Intelligent Metasurface Enabled Multi-User MISO Systems}
\vspace{-.1in}

\author{\IEEEauthorblockN{Qurrat-Ul-Ain Nadeem\IEEEauthorrefmark{1}, Jiancheng An\IEEEauthorrefmark{2}, and Anas Chaaban\IEEEauthorrefmark{1}}
\IEEEauthorblockA{\IEEEauthorrefmark{1} School of Engineering, University of British Columbia, Kelowna, Canada. \\
Email: \{qurrat.nadeem,anas.chaaban\}@ubc.ca}
\IEEEauthorblockA{\IEEEauthorrefmark{2} Engineering Product Development Pillar, Singapore University of Technology and Design, Singapore. \\ Email: jiancheng\_an@sutd.edu.sg}}

\maketitle

\vspace{-.4in}
\begin{abstract}

Stacked intelligent metasurface (SIM) is an emerging programmable metasurface architecture that can implement signal processing directly in the  electromagnetic wave domain, thereby enabling  efficient implementation of ultra-massive multiple-input multiple-output (MIMO) transceivers with a limited number of radio frequency (RF) chains. Channel estimation  (CE) is challenging for SIM-enabled communication systems due to the multi-layer architecture of  SIM, and because we need to estimate large dimensional channels between the SIM  and users with a limited number of RF chains. To efficiently solve this problem, we  develop a novel hybrid digital-wave domain channel estimator, in which the received training symbols  are first processed in the wave domain within the SIM layers, and then processed in the digital domain. The wave domain channel estimator, parametrized by the phase shifts applied by  the meta-atoms in all layers, is optimized  to minimize the mean squared error (MSE)  using a gradient descent algorithm, within which the digital  part is optimally updated. For an SIM-enabled multi-user  system equipped with $4$ RF chains and a $6$-layer SIM with $64$ meta-atoms each, the proposed estimator yields an MSE that is very close to that achieved by fully digital CE in a massive MIMO system employing $64$ RF chains. This high CE accuracy is achieved at the cost of a  training overhead that can   be reduced by  exploiting the potential low rank  of channel correlation matrices. 

\end{abstract}

\vspace{-.1in}
\section{Introduction}
\label{Sec:Intro}


Programmable metasurface has emerged as a key enabler for several potential sixth-generation (6G) communication technologies, including  reconfigurable intelligent surface (RIS) and holographic multiple-input multiple-output (MIMO) communication \cite{SRE1, emimo, HMIMO}. RIS utilizes a large number of passive reflecting elements to  introduce phase shifts onto the incoming electromagnetic (EM) waves, thereby  controlling their propagation in the radio environment to realize desired communication and sensing objectives  \cite{huang, annie}.  However, the promised  gains rely heavily on the availability of accurate channel state information (CSI), that is usually acquired with excessive pilot training overhead \cite{annie_OJ}.  Additionally, the two-hop multiplicative path loss severely impacts the performance of RIS-assisted systems. Placing RISs closer to  communication end-points is often considered to mitigate this loss  \cite{PL}. 

\begin{figure}[t!]
\centering
\includegraphics[scale=.55]{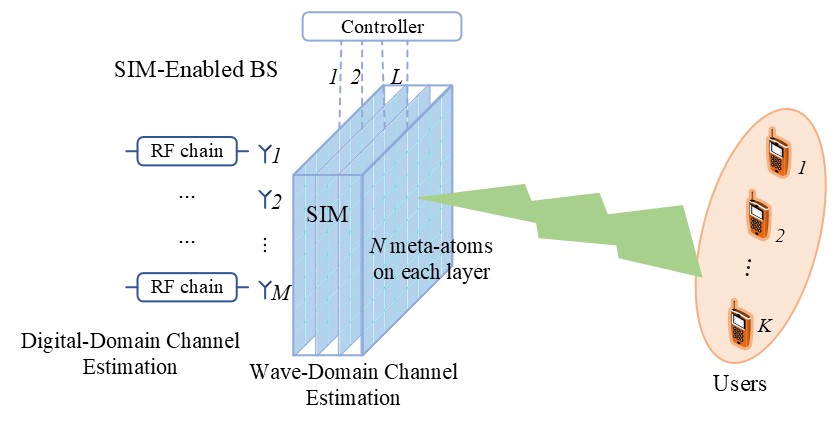}
\caption{SIM-enabled communication system and CE scheme. }
\label{model}
\end{figure}

While  utilizing programmable metasurfaces as reflecting devices (i.e. RISs) in the propagation environment has gained tremendous attention in the last few years, the integration of these surfaces into the  transceiver equipment  is also gaining attention now as an efficient implementation of ultra-massive MIMO and holographic MIMO (HMIMO)  communication systems  \cite{HMIMO}. The conventional implementations of these systems integrate  massive number of active elements at the transceiver and perform signal processing in the  digital domain which  imposes large hardware and computational  costs. Programmable metasurfaces offer low-cost energy-efficient alternative implementations for these systems, and their beam tailoring capabilities can be exploited to realize signal processing in the native analog/EM wave domain \cite{emimo2}.  However the number of meta-atoms that can be integrated in a single-layer metasurface is limited, which restricts its ability to implement beamforming effectively in the wave domain.   To address this limitation,  a stacked intelligent metasurface (SIM)-enabled base station (BS) architecture shown in Fig. \ref{model} has emerged in  \cite{SIM1}, in which multiple metasurface layers are integrated with the conventional radio transceiver that employs a small number of active antennas. The proposed SIM design is inspired by the  wave-based multi-layer  computing paradigm from \cite{DNN}, and utilizes the multiple stacked metasurface layers to offer enhanced signal processing capabilities in the wave domain as  compared to its single-layer counterpart.

The authors in \cite{SIM1} integrated SIMs with the transceivers to support point-to-point HMIMO communication. Under optimized phase-shifts associated with the meta-atoms in all layers, the transmit/receive SIMs are able to implement  precoding/combining automatically as the EM waves propagate through them to  form multiple parallel  sub-channels without any digital precoding/combining, thereby reducing the need for a large number of antennas.   The authors in \cite{SIM2} deployed an SIM-enabled transceiver to perform transmit beamforming directly in the EM wave domain with the performance objective of sum rate maximization. The proposed architecture only requires low-resolution data converters and a moderate number of RF chains that independently process individual users' data streams. Both works  assume the availability of perfect CSI to design the response of the SIMs. 

Channel estimation (CE) in an  SIM-enabled communication system is challenging because (i) the multiple layer structure makes the CE problem  different from those for HMIMO and single/double RIS assisted systems, and (ii) we need to estimate large dimensional channels using a limited number of  RF chains. To  solve this problem, we develop a  hybrid digital-wave domain channel estimator, in which the received training symbols over multiple training sub-phases are first processed in the wave domain within the SIM layers, and then processed in the digital domain. The wave-domain channel estimator, characterized by the phase shifts applied by  the meta-atoms, is optimized  to minimize the mean squared error (MSE)  using a customized gradient descent algorithm, within which the digital  estimator  is optimally computed.  For an SIM-enabled BS employing $4$ RF chains and a $6$-layer SIM with $64$ meta-atoms each, the proposed estimator yields an MSE that approaches that achieved by digital domain CE in a  massive MIMO system  with $64$ RF chains. This high CE accuracy is achieved at the cost of a training overhead that  can be reduced by exploiting the rank structure of the  correlation matrices. Moreover, the multilayer SIM is observed to achieve a target MSE level with a lower  overhead than its single layer counterpart. In this work, we focus on the design of the CE scheme and leave the optimization of its training overhead for future work. We start by describing the system model.  

%

\section{System Model}
\label{Sec:Sys}


We consider an SIM-enabled multi-user multiple-input single-output (MISO)  communication system shown in Fig. \ref{model}, where the BS is equipped with $M$ antennas and an SIM composed of $L$ metasurface layers with $N$ meta-atoms each, while each of the $K$ users has a single antenna. Each meta-atom of the SIM is capable of introducing an independent  phase shift onto the EM waves transmitted through it as directed by a controller. By properly adjusting the phase shifts at each  layer, the SIM can implement precoding/combining directly in the EM wave domain  as discussed in \cite{SIM2}. 

The  phase shifts matrix for the $l$-th layer is denoted as $\boldsymbol{\Theta}^l=\text{diag}(\exp(j\theta^{l}_1), \dots, \exp(j\theta^{l}_N)) \in \mathbb{C}^{N \times N}$, where  $\theta_{n}^l$ is the phase shift introduced by the $n$-th meta-atom in this layer. Further let  $\mathbf{W}^l$ denote the  matrix of transmission coefficients between   $(l-1)$-th  and  $l$-th SIM  layers, with its $(n, n')$-th entry  given according to  Rayleigh-Sommerfeld diffraction theory as \cite{SIM1} \vspace{-.16in}
\begin{align}
\label{W}
& w^l_{n,n'}=\frac{d_1 d_2 \cos \chi_{n,n'}^l}{d_{n,n'}^l}\left(\frac{1}{2\pi d_{n,n'}^l}-\frac{j}{\lambda}  \right) e^{\frac{j 2\pi d_{n,n'}^l}{\lambda}}
\end{align}\normalsize
for $l=2,\dots, L$, where  $\lambda$ is the wavelength, $d_1$ and $d_2$ are the dimensions of each meta-atom, $d_{n,n'}^l$ is the transmission distance, and $\chi^{l}_{n,n'}$ is the angle between the propagation direction and normal direction of the $(l-1)$-th metasurface layer. Similarly define $\mathbf{W}^1 \in \mathbb{C}^{N\times M}$ as the matrix of transmission coefficients between the $M$ transmit antennas and the first metasurface layer of the SIM. The $(n, m)$-th entry of $\mathbf{W}^1$ can be found using \eqref{W} as well. The SIM-enabled wave domain beamforming matrix $\mathbf{G}$ can now be written as
\begin{align}
\label{G}
& \mathbf{G}=\boldsymbol{\Theta}^{L} \mathbf{W}^L \boldsymbol{\Theta}^{L-1} \dots \boldsymbol{\Theta}^2 \mathbf{W}^2  \boldsymbol{\Theta}^1  \in \mathbb{C}^{N\times N}
\end{align}
Note  that the inter-layer transmission coefficients may deviate from \eqref{W} due to  hardware imperfections, and can be  calibrated before deploying the SIM as discussed in \cite{SIM1, SIM2}.

We model the channel $\mathbf{h}_k \in \mathbb{C}^{N\times 1}$ from the last SIM layer to the $k$-th user using correlated Rayleigh fading, i.e., $\mathbf{h}_k \sim \mathcal{CN}(\mathbf{0}, \beta_k \mathbf{R}_k)$,  where $\beta_k$ represents the path loss and $\mathbf{R}_k\in \mathbb{C}^{N\times N}$ is the covariance matrix. By assuming an isotropic scattering environment with uniformly distributed multi-path components, the $(n, n')$-th entry of $\mathbf{R}_k$ is given by \cite{HMIMO3}
\begin{align}
\label{corr}
&[\mathbf{R}_k]_{n,n'} = \text{sinc}(2d_{n,n'}/\lambda),
\end{align}
 where $\text{sinc}(x) = \sin (\pi x) / ( \pi x)$ is the normalized sinc function, and $d_{n,n'}$ denotes the spacing between the meta-atoms.

In sharp contrast to conventional digital precoding schemes that assign each symbol to an individual beamforming vector and perform well when $M\gg K$, SIM-enabled massive MIMO allows for the implementation of precoding/combining  in the wave domain. As a result the users' data symbols are directly transmitted from the antennas, and we can reduce the number of RF chains to as low as $K$ (opposed to $M$ in the conventional case), resulting in huge reduction in cost and power consumption. Existing works exploiting SIM-enabled wave domain beamforming to remove the need for digital precoding  \cite{SIM1}, \cite{SIM2} assume perfect CSI of $\mathbf{h}_k$ to be available  to design  SIM response, which is  unlikely to be true in practice. 

Therefore, our goal in this work is to develop a CE framework to estimate the large $N$ dimensional channel vectors  $\mathbf{h}_k$'s  at the BS that has a limited ($M$) number of RF chains, which makes  CE  challenging. To efficiently solve this problem, we develop a hybrid digital-wave domain channel estimator, in which the  training symbols received over multiple training sub-phases are first processed in the wave domain within the SIM layers, and then processed in the digital domain as shown in Fig. 1,  with the objective of  minimizing the MSE. It should be noted that the phase-shifts of the SIM layers are coupled with inter-layer transmission coefficients which further makes the CE problem  different from those for  HMIMO and RIS-assisted systems \cite{annie_OJ, HMIMO3}.


\section{Channel Estimation}
\label{Sec:Sys2}

In this section we outline and solve the CE problem for the SIM-enabled multi-user MISO system.

\subsection{Uplink Training}

We divide each coherence block into CE and data transmission phases. Throughout the CE phase, the users transmit mutually orthogonal pilot symbols and the BS exploits channel reciprocity under the time division duplex  protocol to estimate the downlink channels using the received uplink pilot signals. Specifically  user $k$ transmits the pilot sequence $\mathbf{x}_{p,k} \in \mathbb{C}^{\tau_p \times 1}$, such that $||\mathbf{x}_{p,k}||^2=\rho_p \tau_p$ and $\mathbf{x}_{p,k}^H \mathbf{x}_{p,k'}=0$ for $k\neq k'$, where $\rho_p=\frac{P_p}{\sigma^2}$ is the training SNR, with $P_p$ being the power of pilot symbol and $\sigma^2$ being the noise variance. To enable the estimation of a large dimensional channel with a limited number of RF chains, we consider $S$ CE sub-phases each of length $\tau_p$ symbols, resulting in a training overhead of $S\tau_p$ symbols. Discussion on the required value of $S$ as a function of $N$, $M$ and $L$ will be provided in Sec. III-C. 

The received training signal at the BS in sub-phase $s$, denoted as $\mathbf{Y}_s \in \mathbb{C}^{M\times \tau_p}$, is given as
\begin{align}
&\mathbf{Y}_s=\sum_{k=1}^K \mathbf{W}^{1^H} \mathbf{G}_s^H \mathbf{h}_k \mathbf{x}_{p,k}^H +\mathbf{N}_s, \hspace{.08in} s=1,\dots, S
\end{align}
where $\mathbf{G}_s=\boldsymbol{\Theta}_s^{L} \mathbf{W}^L \boldsymbol{\Theta}_s^{L-1} \dots \boldsymbol{\Theta}_s^2 \mathbf{W}^2 \boldsymbol{\Theta}_s^1  \in \mathbb{C}^{N \times N}$ is the  wave domain response of the SIM in sub-phase $s$ with $\boldsymbol{\Theta}_s^l=\text{diag}(\exp(j\theta_{s,1}^l), \dots,\exp(j\theta_{s,N}^l) )$, and $\mathbf{N}_s$ represents the noise with each column distributed independently as $\mathcal{CN}(\mathbf{0}, \mathbf{I}_M)$. The BS correlates $\mathbf{Y}_s$ with   pilot sequence of user $k$ to obtain the observation vector for user $k$ in sub-phase $s$ as \vspace{-.12in}
\begin{align}
\label{r}
&\mathbf{r}_{s,k}=\mathbf{W}^{1^H}\mathbf{G}_s^H \mathbf{h}_k + \mathbf{n}_{s,k}/\rho_p \tau_p,  \hspace{.08in} k=1,\dots, K
\end{align}
where $\mathbf{n}_{s,k} =\mathbf{N}_s\mathbf{x}_{p,k}\sim \mathcal{CN}(\mathbf{0},\rho_p \tau_p \mathbf{I}_M)$. Define $\mathbf{r}_k=[\mathbf{r}_{1,k}^T,\dots, \mathbf{r}_{S,k}^T]^T\in \mathbb{C}^{MS\times 1}$, $\mathbf{n}_k=[\mathbf{n}_{1,k}^T,\dots, \mathbf{n}_{S,k}^T]^T\in \mathbb{C}^{MS\times 1}$, $\widetilde{\mathbf{G}}^H=[\mathbf{G}_1 \dots \mathbf{G}_S]^H \in \mathbb{C}^{NS\times N}$, and $\overline{\mathbf{W}}^{1^H} \in \mathbb{C}^{MS\times NS}$  as a block diagonal matrix with each  matrix on the diagonal being $\mathbf{W}^{1^H}$. Collecting the observation vectors in \eqref{r} over the $S$ CE sub-phases we obtain
\begin{align}
\label{rk}
&\mathbf{r}_k=\overline{\mathbf{W}}_1^H \widetilde{\mathbf{G}}^H \mathbf{h}_k + \mathbf{n}_{k}/\rho_p \tau_p,  \hspace{.1in} k=1,\dots, K
\end{align}
where we refer to $\widetilde{\mathbf{G}}$ as the wave-domain channel estimator. Finally, the BS utilizes the digital channel estimator $\mathbf{D}_k \in \mathbb{C}^{MS \times N}$ to estimate $\mathbf{h}_k \in \mathbb{C}^{N \times 1}$ from $\mathbf{r}_{k} \in \mathbb{C}^{MS\times 1}$ as 
\begin{align}
\label{hest}
&\hat{\mathbf{h}}_k=\mathbf{D}_k^H \mathbf{r}_k=\mathbf{D}_k^H \overline{\mathbf{W}}_1^H \widetilde{\mathbf{G}}^H \mathbf{h}_k +  \mathbf{D}_k^H\mathbf{n}_{k}/\rho_p \tau_p
\end{align}


\subsection{Hybrid Digital-Wave Channel Estimator Design}

For the outlined framework, we define the MSE and normalized MSE (NMSE) as 
\begin{align}
\label{MSEo}
&\Gamma_k=\text{tr}(\mathbb{E}[(\hat{\mathbf{h}}_k-\mathbf{h}_k)(\hat{\mathbf{h}}_k-\mathbf{h}_k)^H]) \\
\label{NMSE}
&\overline{\Gamma}_k=\frac{\Gamma_k}{\text{tr}(\mathbb{E}[\mathbf{h}_k \mathbf{h}_k^H ])}
\end{align}
and the average NMSE of the $K$ users as $\overline{\Gamma}=\frac{1}{K}\sum_{k=1}^K \overline{\Gamma}_k$.

Substituting the expression of $\hat{\mathbf{h}}_k$ from \eqref{hest} in \eqref{NMSE}, we can obtain the NMSE in a closed-form as 
\begin{align}
\label{MSE}
\overline{\Gamma}_k&=\frac{1}{\text{tr}(\mathbf{R}_k)}\text{tr}\Big(\mathbf{D}_k^H \overline{\mathbf{W}}^{1^H} \widetilde{\mathbf{G}}^H \mathbf{R}_k \widetilde{\mathbf{G}}\overline{\mathbf{W}}^1 \mathbf{D}_k+\mathbf{R}_k \nonumber \\
& \hspace{.06in}-\mathbf{D}_k^H \overline{\mathbf{W}}^{1^H} \widetilde{\mathbf{G}}^H \mathbf{R}_k-\mathbf{R}_k \widetilde{\mathbf{G}} \overline{\mathbf{W}}^1 \mathbf{D}_k+\frac{\mathbf{D}_k^H\mathbf{D}_k}{\rho_p \tau_p}\Big), \\
&=\frac{1}{\text{tr}(\mathbf{R}_k)} \left(\Big| \Big|(\mathbf{D}_k^H \overline{\mathbf{W}}^{1^H}\hspace{-.06in} \widetilde{\mathbf{G}}^H -\mathbf{I}_N)\sqrt{\mathbf{R}_k}\Big| \Big|^2_{F} + \frac{\text{tr }\mathbf{D}_k^H\mathbf{D}_k}{\rho_p \tau_p}\right) \nonumber
\end{align}

Our objective  is to design the digital-domain part $\mathbf{D}_k$ and the wave-domain part $\widetilde{\mathbf{G}}$, parametrized by $\boldsymbol{\theta}=[\boldsymbol{\theta}^{1^T}_{1}, \dots, \boldsymbol{\theta}^{1^T}_{S}, \boldsymbol{\theta}^{2^T}_{1}, \dots, \boldsymbol{\theta}^{2^T}_{S}, \dots, \boldsymbol{\theta}^{L^T}_{S}]^T\in \mathbb{C}^{NSL\times 1}$ where $\boldsymbol{\theta}^l_s=[\theta^l_{s,1}, \dots, \theta^l_{s,N}]^T \in \mathbb{C}^{N\times 1}$,  of the channel estimator by solving the following NMSE minimization problem.
\begin{subequations}
 \begin{alignat}{2} \textit{(P1)} \hspace{.35in}
&\!\min_{\{\mathbf{D}_k\}_{k=1}^K, \boldsymbol{\theta} }         &\qquad& \overline{\Gamma}  \label{MSEobj}\\
&\text{subject to} &      & \theta^l_{s,n} \in [0, 2\pi ), \hspace{.05in} \forall n,s,l
\end{alignat}
\end{subequations}

 In general, the problem \textit{(P1)} is non-convex as the variables are coupled in the objective function. However for fixed $\tilde{\mathbf{G}}$, the problem can be solved optimally to find $\mathbf{D}_k$, $k=1,\dots, K$ that minimizes the MSE.  Specifically, the optimal $\mathbf{D}_k$ is obtained by solving $\frac{\partial \overline{\Gamma}_k}{\partial \mathbf{D}_k}=0$ to yield 
\begin{align}
\label{D}
&\mathbf{D}_k^*=\left(\overline{\mathbf{W}}^{1^H} \widetilde{\mathbf{G}}^H \mathbf{R}_k \widetilde{\mathbf{G}} \overline{\mathbf{W}}^1+\frac{ \mathbf{I}_{MS} }{\rho_p\tau_p} \right)^{-1} \overline{\mathbf{W}}^{1^H} \widetilde{\mathbf{G}}^H \mathbf{R}_k
\end{align}

However for a given digital part of the channel estimator, it is non-trivial to optimally solve \textit{(P1)} to obtain $\boldsymbol{\theta}^*$ that parametrizes the SIM-enabled wave domain estimator $\widetilde{\mathbf{G}}$ due to the non-convex objective function in which $\boldsymbol{\theta}$ is coupled with  inter-layer transmission coefficients.  We therefore  customize an efficient gradient descent algorithm to solve \textit{(P1)}. 

First the partial derivative of the average NMSE, i.e. the objective  in \eqref{MSEobj}, with respect to the $n$-th phase shift of  $l$-th metasurface layer in  $s$-th training sub-phase is  derived as
\begin{align}
\label{der}
\frac{\partial \bar{\Gamma}}{\partial \theta^l_{s,n}} &=-\frac{2}{K } \sum_{k=1}^K \frac{1}{\text{tr}(\mathbf{R}_k)} \sum_{r=1}^N \sum_{r'=1}^N \Im \Big[ (f_{k,r,r'} -\bar{f}_{k,r,r'}) \nonumber \\
&\hspace{.18in}\times \phi^{l}_{s,n} x^{l^*}_{s,n,k,r,r'}   \Big]
\end{align}
where $f_{k,r,r'}$ is the $(r,r')$-th entry of $\mathbf{F}_{k}=\mathbf{D}_k^H \overline{\mathbf{W}}^{1^H} \widetilde{\mathbf{G}}^H \sqrt{\mathbf{R}_k}$ and $\bar{f}_{k,r,r'}$ is the $(r,r')$-th entry of $\bar{\mathbf{F}}_k=\sqrt{\mathbf{R}_k}$, $\phi^{l}_{s,n}=\exp(j\theta^l_{s,n})$, and $x^l_{s,n,k,r,r'}$ is given by
\begin{align}
\label{x}
&x^l_{s,n,k,r,r'}=[\mathbf{D}_k^H]_{r,:} [\overline{\mathbf{W}}_1^H]_{:,[(s-1)N+1:sN]} \boldsymbol{\Theta}_s^{1^H} \mathbf{W}^{2^H} \dots \nonumber \\
& \dots [\mathbf{W}^{l^H}]_{:,n} [\mathbf{W}^{l+1^H}]_{n,:} \dots \mathbf{W}^{L^H} \boldsymbol{\Theta}^{L^H}_s [\sqrt{\mathbf{R}_k}]_{:,r'},
\end{align}
where $[\mathbf{M}]_{:,[a:b]}$ represents the matrix constructed by extracting $a$-to-$b$-th columns of   $\mathbf{M}$, and $[\mathbf{M}]_{a,:}$ and $[\mathbf{M}]_{:,b}$ represent the $a$-th row vector and $b$-th column vector    of  $\mathbf{M}$ respectively.

Next to avoid potential gradient explosion and  vanishing problems \cite{SIM1}, we normalize the partial derivatives  as 
\begin{align}
\label{norm}
&\frac{\partial \bar{\Gamma}}{\partial \theta^l_{s,n}} \leftarrow \frac{\pi}{\varrho^l_s} \frac{\partial \bar{\Gamma}}{\partial \theta^l_{s,n}} 
\end{align}
where $\varrho^l_s=\underset{n=1,\dots, N}{\text{max}} \left(\frac{\partial \bar{\Gamma}}{\partial \theta^l_{s,n}}  \right)$ denotes the maximum partial derivative value associated with  $l$-th  layer in $s$-th sub-phase. 

Next the phase-shifts at the SIM are updated as 
\begin{align}
\label{update}
\theta^l_{s,n}\leftarrow \theta^l_{s,n} - \eta \frac{\partial \bar{\Gamma}}{\partial \theta^l_{s,n}}
\end{align}
where $\eta$ is the learning rate that determines the step size at each iteration. After the phase-shifts are updated according to \eqref{update} and are used to construct $\widetilde{\mathbf{G}}$  in the iteration, the digital part of the channel estimator, i.e. $\mathbf{D}_k$, $k=1,\dots, K$ is updated using \eqref{D}. Finally the learning rate is updated as 
\begin{align}
\label{eta}
\eta\leftarrow  \beta \eta
\end{align}
 where $0<\beta<1$ is the parameter controlling the decay rate to avoid any overshooting effects.

After repeating \eqref{der} to \eqref{eta} several times, the value of objective function in \eqref{MSEobj} approaches convergence. The overall algorithm to design the hybrid digital-wave domain channel estimator is outlined in Algorithm \ref{Alg1}, and has a polynomial complexity in $M$, $N$, $L$ and $K$ \cite{SIM1}. In order to prevent the algorithm from getting trapped in a local optimum, we generate multiple sets of phase shifts randomly and  select the one that minimizes \eqref{MSEobj} for initialization.  Note that the  parameters of the proposed estimator are functions of only the  correlation matrices and the  inter-layer transmission coefficients, and therefore only need to be updated when the correlation matrices change,  which happens much slower than the fast fading process. Therefore once the estimator is designed, it can be used to compute the estimate $\hat{\mathbf{h}}_k$ in \eqref{hest} in each coherence block  over a set of many blocks. The proposed   estimator will be shown to yield high CE accuracy in the simulations due to the gain provided by the optimization of the wave domain component.  However this accuracy comes at the cost of a training overhead $S$ that is discussed next.  

\begin{algorithm}[!t]
\caption{Gradient Descent Algorithm for Solving \textit{(P1)}}\label{Alg1}
\begin{algorithmic}[1]
\State \textbf{Initialize:} $\mathbf{W}^l$ $l=1\dots L$, $\eta$, $\beta$ and $\epsilon>0$;
\State  \textbf{Initialize:} Phase shifts  $\theta^l_{s,n} \in [0, 2\pi)$ randomly to generate $\boldsymbol{\theta}^{(1)}=[\boldsymbol{\theta}^{1^T}_{1}, \dots, \boldsymbol{\theta}^{L^T}_{S}]^T \in \mathbb{C}^{NSL\times 1}$;
\State \textbf{Construct} $\widetilde{\mathbf{G}}$ and calculate $\mathbf{D}_k^{(1)}$ using \eqref{D};
\State \textbf{Compute} $\bar{\Gamma}^{(1)}$, i.e., the objective in \eqref{MSEobj} and set $p=1$;
\State  \textbf{Repeat}
\State \textbf{Compute} $\frac{\partial \bar{\Gamma}^{(p)}}{\partial \theta^l_{s,n}}$ using \eqref{der};
\State \textbf{Normalize} $\frac{\partial \bar{\Gamma}^{(p)}}{\partial \theta^l_{s,n}}$ using \eqref{norm};
\State \textbf{Update} $\theta^{l^{(p+1)}}_{s,n}\leftarrow \theta^{l^{(p)}}_{s,n} -\eta \frac{\partial \bar{\Gamma}^{(p)}}{\partial \theta^l_{s,n}}$;
\State \textbf{Compute} $\mathbf{D}_k^{(p+1)}$ using \eqref{D};
\State \textbf{Update} $\eta=\beta \eta$;
\State \textbf{Compute} $\bar{\Gamma}^{(p+1)}$ using \eqref{MSEobj} and {update} $p=p+1$;
\State \textbf{Until} $||\bar{\Gamma}^{(p)}-\bar{\Gamma}^{(p-1)}||^2< \epsilon$;
\State \textbf{Obtain} $\boldsymbol{\theta}^*=\boldsymbol{\theta}^{(p)}$ and $\mathbf{D}_k^*=\mathbf{D}_k^{(p)}$, $\forall k$;
\end{algorithmic}
\end{algorithm}


\subsection{How Many Training Sub-phases Are Needed?}

To enable the estimation of a high-dimensional channel using a limited number of RF chains we resort to exploiting multiple ($S$) CE sub-phases as well as optimizing the response of the SIM in the wave domain. The required value of $S$ depends on the values of $N$, $M$ and $L$, the training SNR, the rank of correlation matrices, as well as the tolerable level of NMSE. Given that there is a trade-off between the   durations of CE and  data transmission phases, an important research direction left for future work is to solve a net throughput maximization problem to find the optimal number of training sub-phases that balances the gain from improved CE accuracy and  the rate loss due to time spent in channel training. 

In the scenario where the training SNR is very high, we can infer the number of  sub-phases needed to perfectly estimate $\mathbf{h}_k \in \mathbb{C}^{N\times 1}$ by writing the  observation vectors in \eqref{rk} as
 \begin{align}
 \label{rk1}
&\underset{\rho_p \rightarrow \infty}{\text{lim}} \mathbf{r}_k=\overline{\mathbf{W}}_1^H \widetilde{\mathbf{G}}^H \mathbf{h}_k
 \end{align}

The number of unknown variables and linear equations in \eqref{rk1} are $N$ and $MS$ respectively. As a result, there exists a unique solution to \eqref{rk1} only if the number of linearly independent equations is no smaller than that of the variables, i.e., $S\geq \lceil \frac{N}{M}  \rceil$ . Therefore we need $ S=\lceil \frac{N}{M}  \rceil$ sub-phases, where the SIM adopts different sets of phase shifts in different sub-phases, to perfectly estimate  $\mathbf{h}_k$ with only $M$ RF chains when the training SNR is high. We can also expect, and later see in the simulations, that $S= \lceil \frac{N}{M}  \rceil $ is a reasonable value to accurately estimate $\mathbf{h}_k$ for any practical value of training SNR. As expected, the  required number of sub-phases   increases with the number of meta-atoms in the last layer and decreases with the number of RF chains.  Further, we will  observe in the simulations that the required number of sub-phases to achieve a given NMSE performance decreases with $L$ due to the gain from the  wave-domain component of the channel estimator which, under  optimized  phase shifts, increases with $L$. However, the gain from increasing $L$ eventually saturates due to the fixed thickness of SIM within which deploying excessively dense layers may lead to a performance penalty due to increased correlation and mutual coupling \cite{SIM1}. 




Since $N$ is envisioned  to be large and $M$ to be limited for SIM-enabled  systems, it is  important to develop solutions to reduce the training overhead associated with acquiring the required CSI. One direction to reduce the overhead is to optimize $S$ to balance the  gains from improved CE accuracy and the rate loss due to time spent in training as discussed earlier, resulting in values of $S$ that could be notably less than $\lceil \frac{N}{M}  \rceil$ when net sum rate maximization is the desired objective.  

 Another direction is to exploit the potential low-rank structure of the correlation matrices  to reduce $S$. Note that the rank of  $\mathbf{R}_k$ under isotropic scattering is approximately $R=\pi N \left(\frac{\Delta}{\lambda}\right)^2$ when $N$ is large and inter-element spacing $\Delta$ is small \cite{HMIMO3}, resulting in $80$\% of the eigenvalues of $\mathbf{R}_k$ to be $0$ when $\Delta=\lambda/4$. This observation can be exploited to reduce the  training overhead. To highlight this, we exploit the eigenvalue decomposition of $\mathbf{R}_k$ given as $\mathbf{R}_k=\mathbf{Q}_k \boldsymbol{\Lambda}_k \mathbf{Q}_k^H$, where $\mathbf{Q}_k\in \mathbb{C}^{N\times R}$ and  $\boldsymbol{\Lambda}_k\in \mathbb{C}^{R\times R}$, to write $\mathbf{h}_k$  as 
\begin{align}
\label{low}
&\mathbf{h}_k =\mathbf{Q}_k\tilde{\mathbf{h}}_k
\end{align}
where $\tilde{\mathbf{h}}_k= \boldsymbol{\Lambda}_k^{1/2}  \mathbf{z}_k \in \mathbb{C}^{R\times 1}$ and $\mathbf{z}_k\sim \mathcal{CN}(\mathbf{0},  \mathbf{I}_R)$. The main idea then is to estimate the lower-dimensional channel $\tilde{\mathbf{h}}_k$ using the proposed hybrid digital-wave domain channel estimator and construct $\hat{\mathbf{h}}_k$ using \eqref{low}. For this scenario, the number of CE sub-phases needed for accurate CE  is $S\approx \lceil \frac{R}{M}  \rceil$, which can be significantly less than $S=\lceil \frac{N}{M}  \rceil$. 



\section{Simulation Results}
\label{Sec:Sim}

We consider the SIM layout illustrated  in  \cite[Fig. 2]{SIM1} to test our framework via simulations. Accordingly, the  antenna array is deployed parallel to  $z$-axis and the SIM is deployed parallel to   $x$-$z$ plane, with their centers aligned at a height of $15$ meters from the ground. Each metasurface layer is a uniform planar array with $N_x$ and $N_z$ meta-atoms placed at half-wavelength spacing along the $x$-axis and $z$-axis respectively, where $N = N_xN_z$. We consider the thickness of SIM to be $D_{\rm SIM}=0.05$m, resulting in an inter-layer spacing of $d_l= D_{\rm SIM}/L$.  The dimensions of each meta-atom are $d_1 = d_2 = \lambda/2$.    We consider $K=4$ users located on the $y$-axis, $50$, $60$, $70$ and $80$ meters away from the  SIM, with path loss coefficients $\beta_k$'s computed using \cite[eq (16)]{SIM1}, where $d_0=1$ meter and $b=3.5$. The operating frequency is considered to be $28$GHz and CE parameters are set as $P_p=1$W, $\tau_p=K$,  and  $\sigma^2=-110\rm{dBm}$. In our results,  we compare the  NMSE performance of  following three schemes: 
\begin{enumerate}
\item ``SIM-Optimized" scheme that utilizes the proposed hybrid digital-wave domain estimator outlined in Algorithm 1, with   $\epsilon=10^{-3}$, $\eta=0.1$ and $\beta=0.5$. 
\item ``SIM-Codebook" scheme in which we generate a set of phase shifts vectors $\boldsymbol{\theta}$ randomly, apply the optimal digital estimator in \eqref{D} for each vector, and select the $\boldsymbol{\theta}$ that minimizes $\overline{\Gamma}$. The codebook size is $10LN$. 
\item ``Conventional" massive MIMO system with $M$ RF chains and no SIM, where we utilize the  digital domain MMSE estimation technique \cite{massiveh}. This scheme  is simulated considering the BS has a large number of RF chains  ($M=N$) to  serve as a performance bound.
\end{enumerate}

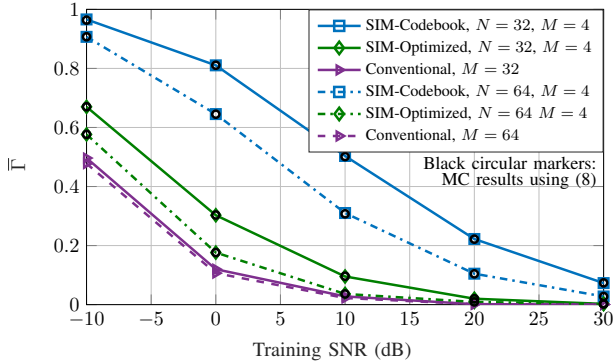
\begin{figure}[!t]
\centering
\tikzset{every picture/.style={scale=.95}, every node/.style={scale=.8}}
%
%
\definecolor{mycolor1}{rgb}{0.00000,0.44706,0.74118}%
\definecolor{mycolor2}{rgb}{0.00000,0.49804,0.00000}%
\definecolor{mycolor3}{rgb}{0.49412,0.18431,0.55686}%
\begin{tikzpicture}

\begin{axis}[%
width=.42\textwidth,
height=.24\textwidth,
scale only axis,
xmin=-10,
xmax=30,
xlabel style={font=\color{white!15!black}},
xlabel={Training SNR (dB)},
ymin=0,
ymax=1,
ylabel style={font=\color{white!15!black}},
ylabel={$\overline{\Gamma}$},
axis background/.style={fill=white},
xmajorgrids,
ymajorgrids,
legend style={at={(axis cs: 30,1)},anchor=north east,legend cell align=left,align=left,draw=white!15!black, /tikz/column 2/.style={
                column sep=5pt,
            }},]
\addplot [color=mycolor1, line width=1.0pt,  mark size=2pt, mark=square, mark options={solid, mycolor1}]
  table[row sep=crcr]{%
-10	0.965943578495346\\
-0	0.810458297688649\\
10	0.503488476669241\\
20	0.22225249057374\\
30	0.0729499856545561\\
40	0.0142475873061058\\
};
\addlegendentry{\footnotesize SIM-Codebook, $N=32$, $M=4$}

\addplot [color=mycolor2, line width=1.0pt, mark size=2.5pt, mark=diamond, mark options={solid, mycolor2}]
  table[row sep=crcr]{%
-10	0.669853300321446\\
0	0.302731890529723\\
10	0.095306411874814\\
20	0.0200006296927955\\
30	0.00246412512547555\\
40	0.000240680823725195\\
};
\addlegendentry{\footnotesize SIM-Optimized, $N=32$, $M=4$}

\addplot [color=mycolor3, line width=1.0pt, mark=triangle, mark options={solid, rotate=270, mycolor3}]
  table[row sep=crcr]{%
-10	0.496402632962876\\
0	0.119904281678254\\
10	0.0278008379765667\\
20	0.00152729403444354\\
30	0.000152730852510553\\
40	1.53192683752452e-05\\
};
\addlegendentry{\footnotesize  Conventional, $M=32$ }

\addplot [color=mycolor1, line width=1.0pt, dashdotted, mark size=2pt, mark=square, mark options={solid, mycolor1}]
  table[row sep=crcr]{%
-10	0.90693738292\\
0	0.6453758483\\
10	0.310258418652952\\
20	0.104747484393\\
30	0.02756373838392\\
40	0.00519483932\\
};
\addlegendentry{\footnotesize  SIM-Codebook, $N=64$, $M=4$}

\addplot [color=mycolor2, line width=1.0pt, dashdotted, mark size=2.5pt, mark=diamond, mark options={solid, mycolor2}]
  table[row sep=crcr]{%
-10	0.576486473827383\\
0	0.176284748838393\\
10	0.036194297999698\\
20	0.009046734627363\\
30	0.002274734838\\
40	0.0005758583\\
};
\addlegendentry{\footnotesize  SIM-Optimized, $N=64$ $M=4$}

\addplot [color=mycolor3, line width=1.0pt, dashed, mark=triangle, mark options={solid, rotate=270, mycolor3}]
  table[row sep=crcr]{%
-10	0.477757818951264\\
0	0.1066120031721\\
10	0.0224441121217813\\
20	0.00151261581002307\\
30	0.000152935625704671\\
40	1.53942326268769e-05\\
};
\addlegendentry{\footnotesize  Conventional, $M=64$}

\addplot [color=black, line width=1.0pt, draw=none, mark=o,mark size=1.5pt, mark options={solid, black}]
  table[row sep=crcr]{%
-10	0.96358319352251\\
0	0.810761492286973\\
10	0.501506436077208\\
20	0.221722191889115\\
30	0.0736784245085896\\
40	0.0142672968684635\\
};

\addplot [color=black, line width=1.0pt, draw=none, mark=o, mark size=1.5pt,mark options={solid, black}]
  table[row sep=crcr]{%
-10	0.670218221336944\\
0	0.303559016915199\\
10	0.096331717845255\\
20	0.0199483638125964\\
30	0.00247620528232897\\
40	0.00024167133905605\\
};

\addplot [color=black, line width=1.0pt, draw=none, mark=o, mark size=1.5pt,mark options={solid, black}]
  table[row sep=crcr]{%
-10	0.90718383939\\
0	0.644938392937\\
10	0.307603517362339\\
20	0.104263637383\\
30	0.02751837483\\
40	0.00538458392\\
};

\addplot [color=black, line width=1.0pt, draw=none, mark=o, mark size=1.5pt,mark options={solid, black}]
  table[row sep=crcr]{%
-10	0.576553526263\\
0	0.1763757383738\\
10	0.03607523708862\\
20	0.00917474837483\\
30	0.0021737382\\
40	0.00056577383\\
};

\node at (axis cs: 30,0.48) [anchor = east] {\small Black circular markers: };
\node at (axis cs: 30,0.42) [anchor = east] {\small  MC results using \eqref{MSEo}};

\end{axis}
\end{tikzpicture}%
\caption{Average NMSE  versus the effective training SNR for $L=6$, $M=K=4$ and $S=\frac{N}{M}$. }
\label{Fig2}
\end{figure}
\begin{figure}[!t]
\centering
\tikzset{every picture/.style={scale=.95}, every node/.style={scale=.8}}
%
%
\definecolor{mycolor1}{rgb}{0.00000,0.44706,0.74118}%
\definecolor{mycolor2}{rgb}{0.00000,0.49804,0.00000}%
\definecolor{mycolor3}{rgb}{0.49412,0.18431,0.55686}%

\begin{tikzpicture}

\begin{axis}[%
width=.4\textwidth,
height=.26\textwidth,
scale only axis,
xmin=1,
xmax=8,
xlabel style={font=\color{white!15!black}},
xlabel={$L$},
ymin=0,
ymax=0.57,
ylabel style={font=\color{white!15!black}},
ylabel={$\overline{\Gamma}$},
axis background/.style={fill=white},
xmajorgrids,
ymajorgrids,
legend style={at={(axis cs: 8,0.57)},anchor=north east,legend cell align=left,align=left,draw=white!15!black, /tikz/column 2/.style={
                column sep=5pt,
            }},]

\addplot [color=mycolor1, line width=1.0pt, mark size=2pt, mark=square, mark options={solid, mycolor1}]
  table[row sep=crcr]{%
1	0.489393416162084\\
2	0.456318016210192\\
4	0.417244607758425\\
6	0.434711361800839\\
8	0.428344268073593\\
};
\addlegendentry{\footnotesize SIM-Codebook}

\addplot [color=mycolor2, line width=1.0pt, mark size=2.5pt, mark=diamond, mark options={solid, mycolor2}]
  table[row sep=crcr]{%
1	0.320123434588763\\
2	0.214625446718158\\
4	0.123709614276233\\
6	0.10067282392298\\
8	0.098384196221412\\
};
\addlegendentry{\footnotesize SIM-Optimized}

%

\addplot [color=mycolor3, dotted, line width=1.0pt, mark size=2pt, mark=triangle, mark options={solid, rotate=270, mycolor3}]
  table[row sep=crcr]{%
1	0.022060074528491\\
2	0.022088074766353\\
4	0.022083738413827\\
6	0.022083738413827\\
8	0.022083738413827\\
};
\addlegendentry{\footnotesize Conventional, $M=64$}

\addplot [color=mycolor1, dashdotted, line width=1.0pt, mark size=2pt, mark=square, mark options={solid, mycolor1}]
  table[row sep=crcr]{%
1	0.348735662879882\\
2	0.284419131767709\\
4	0.233435993122334\\
6	0.245135353555558\\
8	0.243757483844294\\
};

\addplot [color=mycolor2, dashdotted, line width=1.0pt, mark size=2.5pt, mark=diamond, mark options={solid, mycolor2}]
  table[row sep=crcr]{%
1	0.24204063363776\\
2	0.11855507375432\\
4	0.05873909497078\\
6	0.04443242424355\\
8	0.04235373733535\\
};

\addplot [color=black, line width=1.0pt, draw=none, mark size=1.5pt, mark=o, mark options={solid, black}]
  table[row sep=crcr]{%
1	0.489748384938493\\
2	0.456374374932932\\
4	0.418589375938394\\
6	0.431938937498494\\
8	0.429478364837483\\
};

\addplot [color=black, line width=1.0pt, draw=none, mark=o, mark size=1.5pt,mark options={solid, black}]
  table[row sep=crcr]{%
1	0.320813561077192\\
2	0.213443821588059\\
4	0.12421658810743\\
6	0.10067282392298\\
8	0.098384196221412\\
};

\addplot [color=black, line width=1.0pt, draw=none, mark=o, mark size=1.5pt,mark options={solid, black}]
  table[row sep=crcr]{%
1	0.348726374735\\
2	0.285433984938\\
4	0.233435993122\\
6	0.245185453553\\
8	0.243757483894\\
};

\addplot [color=black, line width=1.0pt, draw=none, mark=o, mark size=1.5pt,mark options={solid, black}]
  table[row sep=crcr]{%
1	0.243727458053162\\
2	0.119595839628037\\
4	0.058889019344290\\
6	0.044533353535353\\
8	0.042353737535333\\
};

\node at (axis cs: 8,0.36) [anchor = east] {\small  Solid Lines: $N=32$, $M=4$};
\node at (axis cs: 8,0.32) [anchor = east] {\small  Dashdotted Lines: $N=64$, $M=4$};

\end{axis}
\end{tikzpicture}%
\caption{Average NMSE  versus $L$ for $M\hspace{-.03in}=\hspace{-.03in}K\hspace{-.03in}=4$, and $S=\frac{N}{M}$.}
\label{Fig3}
\end{figure}
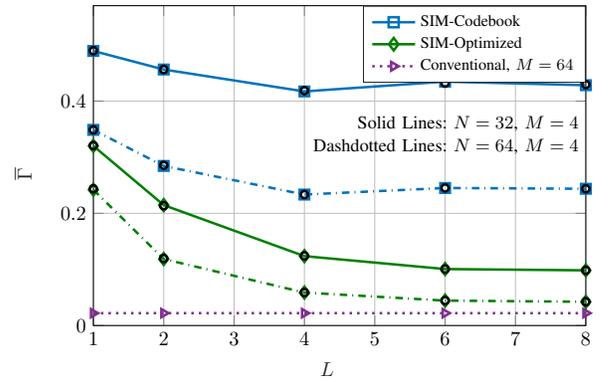

In Fig. \ref{Fig2} we plot   $\overline{\Gamma}=\frac{1}{K} \overline{\Gamma}_k$, where $\overline{\Gamma}_k$ is given by \eqref{NMSE} and can be computed theoretically using \eqref{MSE} and numerically using Monte-Carlo (MC) simulation of \eqref{MSEo} over $1,000$ channel realizations. The results are plotted against the effective training SNR abstracted as $\frac{P_p \bar{\beta}}{\sigma^2}$ where $\bar{\beta}=\frac{1}{K}\sum_{k=1}^K \beta_k$  and $\sigma^2$ is varied to get different training SNRs.  We consider the  SIM to have $L=6$ layers with $N=32$ or $64$ meta-atoms in each layer.  The goal is therefore to estimate the $32$ or $64$-dimensional channel using  $M=4$ RF chains at the BS, utilizing $S=\frac{N}{M}$ sub-phases. The proposed optimized digital-wave domain channel estimator achieves a significantly lower NMSE compared to the codebook-based scheme across all training SNR values. Further we observe that increasing $N$ from $32$ to $64$ results in smaller NMSE even though we are estimating a larger dimensional channel with a fixed number of RF chains.  This is because as $N$ increases, we  utilize a larger number of sub-phases ($S=\frac{N}{M}$), and also optimize a larger number of meta-atoms in each of the intermediate layers of the SIM to improve the performance of the wave-domain estimator resulting in higher CE accuracy. Finally, we  also consider the conventional massive MIMO system that does not employ  SIM and rather implements signal processing entirely in the digital domain. The proposed hybrid channel estimator for an SIM-enabled system that employs only $4$ RF chains and $64$ meta-atoms in each  layer is seen to perform quite close in terms of NMSE to the fully digital CE scheme implemented at a conventional BS using $64$ RF chains. 


Next we plot the average NMSE against the number of SIM layers $L$ in Fig. \ref{Fig3}. The NMSE achieved by the  optimized hybrid digital-wave domain CE scheme decreases with $L$, benefiting from the larger gain that the SIM's wave-domain response yields.  We also note that the NMSE gradually converges as $L$ increases. It reaches the minimum at approximately $L = 6$ layers, achieving a $69$\% improvement in average NMSE compared with a single-layer SIM, when considering $N = 32$ meta-atoms in each layer. Therefore a multi-layer SIM  results in better CE performance under Algorithm 1 compared to its single-layer counterpart.  We also  see that the optimized hybrid channel estimator for an SIM-enabled MISO  system employing $4$ RF chains and a $6$-layer-SIM with  $64$ meta-atoms each achieves NMSE similar to that achieved by the conventional digital estimator for a massive MIMO system employing $64$ RF chains, albeit at the cost of a larger training overhead that is studied next. 
\begin{figure}[!t]
\centering
\tikzset{every picture/.style={scale=.95}, every node/.style={scale=.8}}
%
%
\definecolor{mycolor1}{rgb}{0.00000,0.49804,0.00000}%
\begin{tikzpicture}

\begin{axis}[%
width=.42\textwidth,
height=.2\textwidth,
scale only axis,
xmin=4,
xmax=14,
xlabel style={font=\color{white!15!black}},
xlabel={S},
ymin=0,
ymax=0.64,
ylabel style={font=\color{white!15!black}},
ylabel={$\overline{\Gamma}$},
axis background/.style={fill=white},
xmajorgrids,
ymajorgrids,
legend style={at={(axis cs: 14,0.64)},anchor=north east,legend cell align=left,align=left,draw=white!15!black, /tikz/column 2/.style={
                column sep=5pt,
            }},]

\addplot [color=mycolor1, line width=1.0pt, mark size=2pt, mark=diamond, mark options={solid, mycolor1}]
  table[row sep=crcr]{%
4	0.592435584528839\\
6	0.450869105463435\\
8	0.328139456051973\\
10	0.22522322493995\\
12	0.144644247030479\\
14	0.091587203219459\\
16	0.0625729836793323\\
};
\addlegendentry{\footnotesize SIM-Optimized  $L=2$, $M=4$}
						
\addplot [color=mycolor1, line width=1.0pt, mark size=2.0pt, mark=star, mark options={solid, mycolor1}]
  table[row sep=crcr]{%
4	0.53643735479699\\
6	0.377769364866592\\
8	0.248039223776182\\
10	0.132995505577547\\
12	0.0707510280016205\\
14	0.0525521791391548\\
16	0.0409835845267187\\
};
\addlegendentry{\footnotesize SIM-Optimized  $L=4$, $M=4$}

\addplot [color=mycolor1, line width=1.0pt, mark size=1.6pt, mark=square, mark options={solid, mycolor1}]
  table[row sep=crcr]{%
4	0.507994090787209\\
6	0.352633740012276\\
8	0.221205295404166\\
10	0.110135691318868\\
12	0.0541413729678321\\
14	0.0395813152681972\\
};
\addlegendentry{\footnotesize SIM-Optimized $L=6$, $M=4$}

\addplot [color=mycolor1, dashed, line width=1.0pt, mark size=2pt, mark=diamond, mark options={solid, mycolor1}]
  table[row sep=crcr]{%
4	0.430486467827946\\
6	0.252806473464258\\
8	0.120442471935694\\
10	0.0679848787731811\\
12	0.0495029131916966\\
14	0.0381584115193244\\
};
\addlegendentry{\footnotesize SIM-Optimized $L=2$, $M=8$}

\addplot [color=mycolor1, dashed, line width=1.0pt, mark size=2.0pt, mark=star, mark options={solid, mycolor1}]
  table[row sep=crcr]{%
4	0.343617254925984\\
6	0.147934168964409\\
8	0.0700117776205563\\
10	0.048679457967058\\
12	0.0363864034574229\\
14	0.0283191440328955\\
};
\addlegendentry{\footnotesize SIM-Optimized $L=4$, $M=8$}

\addplot [color=mycolor1, dashdotted, line width=1.0pt, mark size=1.6pt, mark=square, mark options={solid, mycolor1}]
  table[row sep=crcr]{%
4	0.312288456903438\\
6	0.112153428633229\\
8	0.0552784063021093\\
10	0.0416957358146287\\
12	0.0308619348949742\\
14	0.026104680633975\\
};
\addlegendentry{\footnotesize SIM-Optimized $L=6$, $M=8$}

\end{axis}
\end{tikzpicture}%
\caption{Average NMSE versus $S$ for $N=48$ and $K=4$.}
\label{Fig4}
\end{figure}
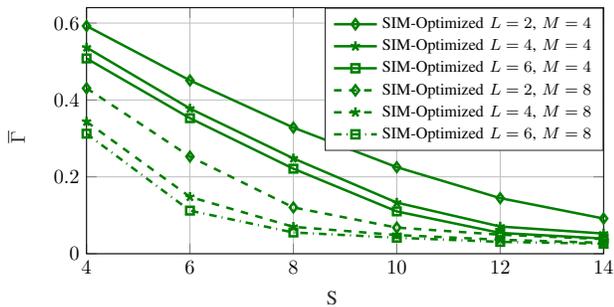

\begin{figure}[!t]
\centering
\tikzset{every picture/.style={scale=.95}, every node/.style={scale=.8}}
%
%
\definecolor{mycolor1}{rgb}{0.00000,0.44706,0.74118}%
\definecolor{mycolor2}{rgb}{0.00000,0.49804,0.00000}%
\begin{tikzpicture}

\begin{axis}[%
width=.42\textwidth,
height=.2\textwidth,
scale only axis,
xmin=0,
xmax=30,
xlabel style={font=\color{white!15!black}},
xlabel={Iteration},
ymin=0.05,
ymax=0.4,
ylabel style={font=\color{white!15!black}},
ylabel={$\overline{\Gamma}$},
axis background/.style={fill=white},
axis x line*=bottom,
axis y line*=left,
xmajorgrids,
ymajorgrids,
legend style={at={(axis cs: 30,0.4)},anchor=north east,legend cell align=left,align=left,draw=white!15!black, /tikz/column 2/.style={
                column sep=5pt,
            }},]

\addplot [color=mycolor2, line width=1.0pt, mark=diamond, mark options={solid, mycolor2}]
  table[row sep=crcr]{%
1	0.363285749493999\\
2	0.326030103933472\\
3	0.299722922416028\\
4	0.280145290595346\\
5	0.265462714322157\\
6	0.25445610203797\\
7	0.246132439967842\\
8	0.239838018132388\\
9	0.235136662670782\\
10	0.231468194832199\\
11	0.228430580967344\\
12	0.225819852045608\\
13	0.223589617110369\\
14	0.221709360982691\\
15	0.220104075617893\\
16	0.218724559073738\\
17	0.217538986998576\\
18	0.216519927635542\\
19	0.215639986405804\\
20	0.214876343696964\\
21	0.214207800074438\\
22	0.213618525032132\\
23	0.213096803311904\\
24	0.21263392694997\\
25	0.212222695783018\\
26	0.211856289596951\\
27	0.211529064436151\\
28	0.211236485459148\\
29	0.210975683832856\\
30	0.210742840698616\\
31	0.210534747301079\\
32	0.210348740188889\\
};
\addlegendentry{\footnotesize SIM-Optimized, $L=2$, $N=32$}

\addplot [color=mycolor1, line width=1.0pt, mark=x, mark options={solid, mycolor1}]
  table[row sep=crcr]{%
1	0.35951748392708\\
2	0.299440389823933\\
3	0.257694115328705\\
4	0.227370736121976\\
5	0.205298921312239\\
6	0.189314824400649\\
7	0.177436236570664\\
8	0.168124942946497\\
9	0.160558537313861\\
10	0.154587997275872\\
11	0.149640557335573\\
12	0.145618510730714\\
13	0.142115838642402\\
14	0.139235094448636\\
15	0.136712833896109\\
16	0.13460951261347\\
17	0.132759615577805\\
18	0.131156918459027\\
19	0.129742459669553\\
20	0.12850745636376\\
21	0.12742418983998\\
22	0.126476801726813\\
23	0.125641320940601\\
24	0.124911361858584\\
25	0.124265620999512\\
26	0.123696754732563\\
27	0.123193513209179\\
28	0.122748751177856\\
29	0.122354630965781\\
30	0.122004405612895\\
31	0.121692540183369\\
32	0.121414279721566\\
33	0.121165780336316\\
34	0.12094370969485\\
35	0.120745032731235\\
36	0.120567120443192\\
37	0.120407680874706\\
};
\addlegendentry{\footnotesize SIM-Optimized, $L=6$, $N=32$}

\addplot [color=mycolor2, dashdotted, line width=1.0pt, mark=diamond, mark options={solid, mycolor2}]
  table[row sep=crcr]{%
1	0.236362949997028\\
2	0.206995561063418\\
3	0.187863262085632\\
4	0.174205028575837\\
5	0.163820681750064\\
6	0.156139146925242\\
7	0.150182529859956\\
8	0.145539548342697\\
9	0.141794079480631\\
10	0.138756374268378\\
11	0.136204447071486\\
12	0.134066251762797\\
13	0.132261887107349\\
14	0.130744099628848\\
15	0.129463397795975\\
16	0.128378696827102\\
17	0.12744878227899\\
18	0.126646631897564\\
19	0.125950926423269\\
20	0.125342840375715\\
21	0.124809304847284\\
22	0.124339482219483\\
23	0.123924127284149\\
24	0.123556082398155\\
25	0.12322977183834\\
26	0.122939492024988\\
27	0.122680724957864\\
28	0.122449640937447\\
29	0.122242875143624\\
30	0.122057628844866\\
31	0.121891603475992\\
32	0.121742842523201\\
33	0.1216096012331\\
34	0.12149019007452\\
35	0.121383079207058\\
36	0.121287007516918\\
37	0.121200811294031\\
38	0.121123446607001\\
};
\addlegendentry{\footnotesize SIM-Optimized, $L=2$, $N=64$}

\addplot [color=mycolor1, dashdotted, line width=1.0pt, mark=x, mark options={solid, mycolor1}]
  table[row sep=crcr]{%
1	0.202473939215221\\
2	0.161914511341878\\
3	0.136357851806018\\
4	0.120026716458245\\
5	0.108752648669766\\
6	0.100600540756128\\
7	0.0941653922745326\\
8	0.0889795320277822\\
9	0.0847640984649954\\
10	0.0813689970529075\\
11	0.078470924321905\\
12	0.0761070039134772\\
13	0.07399913820427\\
14	0.0721832287016315\\
15	0.0705713819549592\\
16	0.0691954844920387\\
17	0.0679820042094513\\
18	0.0669387994645816\\
19	0.0660430208956832\\
20	0.0652581935874609\\
21	0.064577023432959\\
22	0.0639841400961138\\
23	0.0634662972543622\\
24	0.0630117844854025\\
25	0.0626109588724161\\
26	0.0622566382051127\\
27	0.0619430941274119\\
28	0.0616654239424277\\
29	0.061419307642707\\
30	0.0612007561276267\\
31	0.0610065332268342\\
32	0.0608337017930392\\
33	0.0606796679885266\\
34	0.0605422155603404\\
35	0.06041941936174\\
36	0.0603096222494135\\
37	0.0602113833728408\\
38	0.0601234373002534\\
39	0.0600446685925165\\
40	0.059974084606959\\
41	0.059910800621648\\
42	0.0598540351661015\\
};
\addlegendentry{\footnotesize SIM-Optimized, $L=6$, $N=64$}

\end{axis}
\end{tikzpicture}%
\caption{Convergence of Algorithm 1 for  $M,K\hspace{-.04in}=\hspace{-.04in}4$ and $S=\frac{N}{M}$.}
\label{Fig5}
\end{figure}
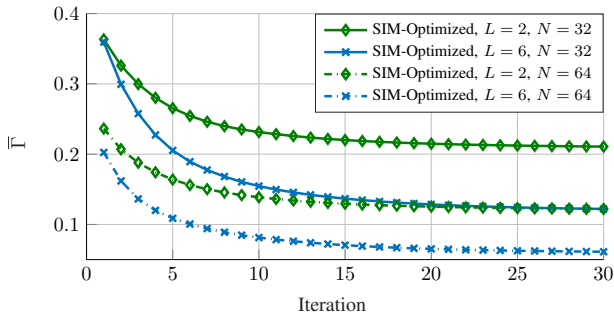

Next we study NMSE performance of the proposed hybrid CE framework against the number of training sub-phases $S$ in Fig. \ref{Fig4}, and observe the NMSE to decease with  $S$ as expected.  We also note that $S=\lceil \frac{N}{M}  \rceil$ sub-phases, i.e. $S=6$ for $M=8$ and $S=12$ for $M=4$, result in low CE error and is a reasonable measure of the number of  sub-phases required to yield accurate CSI.  However, it is not desirable to require large values of $S$ especially in scenarios with short coherence block lengths. One way to decrease the value of $S$ required to achieve a target NMSE threshold is to deploy more RF chains. i.e. increase $M$. Further, we see that an SIM with a larger number of layers requires a smaller value of $S$ to achieve a target NMSE threshold  due to the improved  performance of the wave-domain channel estimator. For example,  when estimating  channels of dimension $N=48$ between the last SIM layer and the users at the BS that has $M=8$ RF chains, the proposed protocol requires around $S=6$ sub-phases for $L=6$, and $S=9$ sub-phases for $L=2$ layers to guarantee an NMSE level of $0.1$. Similarly, when the BS has $M=4$ RF chains, these number are $S=10$ for $L=6$ and $S=14$ for $L=2$ layers.  The proposed work, being the first on CE for SIM-enabled communication systems, provides an important benchmark for the training overhead required to accurately estimate the channels using a limited number of RF chains under pilot based CE schemes. The analysis of  optimal value of $S$ to maximize the net sum rates is left for future work. 


In Fig. \ref{Fig5} we observe that the proposed gradient descent algorithm to design the  hybrid digital-wave domain channel estimator converges after a moderate number of iterations. The average NMSE for the largest considered system ($N=64$ and $L=6$) converges within $30 $ iterations. Note that Algorithm 1 only needs to be implemented when the channel correlation matrices change. Therefore, the designed estimator can be used to estimate channels in multiple coherence blocks over which the channel correlation matrices are unchanged.


\section{Conclusion}
\label{Sec:Con}

In this work, we developed a hybrid digital-wave domain CE framework for the new SIM-enabled multi-user MISO communication paradigm to allow the BS to estimate  high dimensional channels  with a limited number of RF chains.  The  training symbols received over multiple CE sub-phases are first processed in the wave domain within the SIM layers, and then processed in the digital domain. The wave-domain channel estimator is optimized to minimize the NMSE  using a gradient descent algorithm, within which the digital estimator is optimally updated. The proposed hybrid estimator for an SIM-enabled system is shown to yield an NMSE that is very close to that achieved by digital domain CE in a massive MIMO system employing a much larger number of RF chains. The high CE accuracy comes at the cost of a training overhead that can optimized in future works or reduced by exploiting the rank deficiency in correlation matrices. Further, we observe the multi-layer SIM to achieve a given NMSE level with a lower  overhead than its single layer counterpart.

\bibliographystyle{IEEEtran}
\bibliography{bib}
\end{document}